\documentclass[12pt,a4paper]{article}	
\usepackage[british]{babel}
\usepackage{epsfig}
\begin{document}
\date{}

\title{
{\vspace{-20mm} \normalsize
\hfill \parbox[t]{50mm}{\small DESY 06-092     \\
                               MS-TP-06-2      \\
                               BNL-HET-06/3}}  \\[25mm]
 Numerical simulation of QCD with $u$, $d$, $s$ and $c$ \\
 quarks in the twisted-mass Wilson formulation \\[5mm]}

\author{T.\ Chiarappa$^{a}$,
        F.\ Farchioni$^{b}$,
        K.\ Jansen$^{c}$,
        I.\ Montvay$^{d}$,\\
        E.E.\ Scholz$^{e}$,
        L.\ Scorzato$^{f}$,
        T.\ Sudmann$^{b}$,
        C.\ Urbach$^{g}$\\[5mm]
 {\small $^a$  Universit\`a Milano Bicocca, Piazza della Scienza 3,
  I-20126 Milano, Italy}\\
 {\small $^b$ Universit\"at M\"unster,
  Institut f\"ur Theoretische Physik,}\\
 {\small Wilhelm-Klemm-Strasse 9, D-48149 M\"unster,
  Germany}\\
 {\small $^c$ NIC, DESY, Zeuthen, Platanenallee 6, D-15738 Zeuthen,
  Germany}\\
 {\small $^d$ Deutsches Elektronen-Synchrotron DESY, Notkestr.\,85,
  D-22603 Hamburg, Germany}\\
 {\small $^e$ Physics Department, Brookhaven National Laboratory,
  Upton, NY 11973 USA}\\
 {\small $^f$ ECT* strada delle tabarelle 286, 38050 Villazzano (TN), Italy}\\
 {\small $^g$ Theoretical Physics Division, Dept. of Mathematical Sciences,}\\
 {\small University of Liverpool, Liverpool L69 3BX, UK}\\[5mm]}
%
\newcommand{\be}{\begin{equation}}
\newcommand{\ee}{\end{equation}}
\newcommand{\bea}{\begin{eqnarray}}
\newcommand{\eea}{\end{eqnarray}}
\newcommand{\half}{\frac{1}{2}}
\newcommand{\rar}{\rightarrow}
\newcommand{\lar}{\leftarrow}
\newcommand{\LCB}{\raisebox{-0.3ex}{\mbox{\LARGE$\left\{\right.$}}}
\newcommand{\RCB}{\raisebox{-0.3ex}{\mbox{\LARGE$\left.\right\}$}}}
\newcommand{\LSB}{\raisebox{-0.3ex}{\mbox{\LARGE$\left[\right.$}}}
\newcommand{\RSB}{\raisebox{-0.3ex}{\mbox{\LARGE$\left.\right]$}}}
\newcommand{\tr}{{\rm Tr}}
\newcommand{\alphah}{\frac{\omega_l}{2}}
\newcommand{\betah}{\frac{\omega_h}{2}}
\newcommand{\cprime}{c_h}
\newcommand{\sprime}{s_h}
\newcommand{\cli}{c_l}
\newcommand{\sli}{s_l}

\maketitle

\abstract{A first study of numerical Monte Carlo simulations with two
 quark doublets, a mass-degenerate one and a mass-split one,
 interpreted as $u$, $d$, $s$ and $c$ quarks, is carried out in the
 framework of the twisted mass Wilson lattice formulation.
 Tuning the bare parameters of this theory is explored on $12^3 \cdot 24$
 and $16^3 \cdot 32$ lattices at lattice spacings
 $a \simeq 0.20 {\rm\,fm}$ and $a \simeq 0.15 {\rm\,fm}$, respectively.}

\newpage
\section{Introduction}\label{sec1}

 In QCD the effect of virtual quark loops is most important for the
 three light quarks $(u,d,s)$.
 In recent unquenched numerical simulations, besides the two lightest
 quarks $u$ and $d$, the $s$-quark is also included (see, for instance,
 \cite{MILC,JLQCD}).
 The formulation of QCD with twisted-mass Wilson fermions \cite{Frezzotti:tmqcd}
 is based on chiral rotations of the bare mass (or, equivalently, of
 the Wilson-term) within quark doublets.
 Therefore, in this formulation there are two possibilities for
 unquenched simulations with $(u,d,s)$ quarks: either the $s$ quark is
 taken alone and the twisted mass formulation is restricted to the
 $(u,d)$ doublet or, in addition to the $s$ quark, also the $c$-quark
 is included in a mass-split doublet using the formulation in 
 Ref.~\cite{FrezzottiRossi:split}.
 In the present paper we explore the latter possibility (for first
 results along this line see the proceedings contribution \cite{LAT05}).

 Numerical simulations with twisted-mass Wilson fermions are usually
 performed at (or near) the critical (untwisted) bare quark mass,
 because there an {\em automatic} ${\cal O}(a)$ {\em improvement of the
 continuum limit} is expected \cite{FREZZOTTI-ROSSI}.
 Our collaboration has performed several studies of twisted-mass QCD
 both in the quenched approximation \cite{Jansen:quenched},
 \cite{Bietenholz:overlapversustm}, \cite{Jansen:quenchedlight},
 \cite{Jansen:quenchedscaling} and in unquenched simulations with
 dynamical $(u,d)$ quarks \cite{Farchioni:tmphases},
 \cite{Farchioni:dbw2}, \cite{Farchioni:latticespacing},
 \cite{Farchioni:dbw2two}.
 In the present paper we explore the possibility of numerical
 simulations of QCD with a degenerate doublet ($u,d$) and an mass-split
 doublet ($c,s$) of dynamical quarks in the twisted mass Wilson
 formulation.

 The plan of this paper is as follows: in the next section we define
 the lattice action and describe the simulation algorithm.
 Section \ref{sec3} is devoted to the introduction of physical fields
 and currents important for the interpretation of results.
 In Section \ref{sec4} we present our numerical results.
 The last section contains a discussion and final remarks.

\section{Lattice action and simulation algorithm}\label{sec2}

\subsection{Lattice action}\label{sec2.1}

 The notation for the lattice action of the {\em light} mass-degenerate
 $(u,d)$-doublet, denoted by a subscript $l$, is the same as in our
 previous papers, for instance, Ref.~\cite{Farchioni:dbw2two}:
\bea\label{eq01}
S_l & = & \sum_x \left\{ 
\left( \overline{\chi}_{l,x} [\mu_{\kappa l} + i\gamma_5\tau_3\,a\mu_l]
\chi_{l,x} \right)
- \half\sum_{\mu=\pm 1}^{\pm 4}
\left( \overline{\chi}_{l,x+\hat{\mu}}U_{x\mu}[r+\gamma_\mu]
\chi_{l,x} \right) \right\} 
\nonumber\\[0.5em]
& \equiv & \sum_{x,y} \overline{\chi}_{l,x} Q^{(\chi)}_{l,xy}
\chi_{l,y} \ .
\eea
 Here, and in most cases below, colour-, spinor- and isospin indices are
 suppressed.
 For the isospin indices later on we shall also use notations as, for
 instance, $\chi_l \equiv (\chi_u\; \chi_d)$.
 The (``untwisted'') bare quark mass of the light doublet in
 lattice units is denoted by
\be\label{eq02}
\mu_{\kappa l} \equiv am_{0l} + 4r = \frac{1}{2\kappa_l} \ ,
\ee
 $r$ is the Wilson-parameter, set in our simulations to $r=1$, $am_{0l}$
 is another convention for the bare quark mass in lattice units and
 $\kappa_l$ is the conventional hopping parameter.
 The twisted mass of the light doublet in lattice units is denoted
 by $a\mu_l$.
 $U_{x\mu} \in {\rm SU(3)}$ is the gauge link variable and we also
 defined $U_{x,-\mu} = U_{x-\hat{\mu},\mu}^\dagger$ and
 $\gamma_{-\mu}=-\gamma_\mu$.

 Besides the quark doublet fields $\chi_l,\overline{\chi}_l$ in
 (\ref{eq01}) it will turn out convenient to introduce other fields
 by the transformation
\be\label{eq03}
\psi_{l,x} \equiv \frac{1}{\sqrt{2}} \left(1+i\gamma_5\tau_3\right)
\chi_{l,x} \ , \hspace*{3em}
\overline{\psi}_{l,x} \equiv \overline{\chi}_{l,x}
\frac{1}{\sqrt{2}} \left(1+i\gamma_5\tau_3\right) \ .
\ee
 The {\em quark matrix on the $\chi$-basis} $Q^{(\chi)}_l$ defined
 in (\ref{eq01}) is
\be\label{eq04}
Q^{(\chi)}_{l,xy} = \delta_{xy} 
\left( \mu_{\kappa l} + i\gamma_5\tau_3\,a\mu_l \right)
-\half \sum_{\mu=\pm 1}^{\pm 4} \delta_{x,y+\hat{\mu}} U_{y\mu}
[r+\gamma_\mu]
\ee
 or in a short notation, without the site indices,
\be\label{eq05}
Q^{(\chi)}_l = \mu_{\kappa l} + i\gamma_5\tau_3\,a\mu_l + N + R \ ,
\ee
 with
\be\label{eq06}
N_{xy} \equiv -\half \sum_{\mu=\pm 1}^{\pm 4} \delta_{x,y+\hat{\mu}} 
U_{y\mu}\gamma_\mu \ , \hspace*{2em}
R_{xy} \equiv -\frac{r}{2} \sum_{\mu=\pm 1}^{\pm 4} 
\delta_{x,y+\hat{\mu}} U_{y\mu} \ .
\ee
 On the {\em $\psi$-basis} defined in (\ref{eq03}) we have the
 quark matrix
\be\label{eq07}
Q^{(\psi)}_l = \half \left(1-i\gamma_5\tau_3\right) Q^{(\chi)}_l
\left(1-i\gamma_5\tau_3\right) =
a\mu_l + N -i\gamma_5\tau_3 \left(\mu_{\kappa l}+R\right) \ .
\ee

 As it has been shown by Frezzotti and Rossi in
 Ref.~\cite{FrezzottiRossi:split}, a {\em real quark determinant} with
 a mass-split doublet can be obtained if the mass splitting is taken to
 be orthogonal in isospin space to the twist direction.
 One could take, for instance, the mass splitting in the $\tau_1$
 direction if the twist is in the $\tau_3$ direction, as in
 (\ref{eq01}).
 It is, however, more natural to exchange the two directions because
 then the bare mass is diagonal in isospin.
 In this case, the lattice action of the {\em heavy} mass-split
 $(c,s)$-doublet, denoted by a subscript $h$, is defined as
\be\label{eq08}
S_h = \sum_{x,y} \overline{\chi}_{h,x} Q^{(\chi)}_{h,xy}
\chi_{h,y}
\ee
 with
\be\label{eq09}
Q^{(\chi)}_h = \mu_{\kappa h} + i\gamma_5\tau_1\,a\mu_\sigma
+ \tau_3 a\mu_\delta + N + R \ .
\ee
 For the isospin indices later on we shall also use notations as, for
 instance, $\chi_h \equiv (\chi_c\; \chi_s)$.
 The {\em $\psi$-basis} is introduced similarly to (\ref{eq03}) by
\be\label{eq10}
\psi_{h,x} \equiv \frac{1}{\sqrt{2}} \left(1+i\gamma_5\tau_1\right)
\chi_{h,x} \ , \hspace*{3em}
\overline{\psi}_{h,x} \equiv \overline{\chi}_{h,x}
\frac{1}{\sqrt{2}} \left(1+i\gamma_5\tau_1\right) 
\ee
 and the {\em quark matrix on the $\psi$-basis} is for the heavy
 mass-split doublet
\be\label{eq11}
Q^{(\psi)}_h = \half \left(1-i\gamma_5\tau_1\right) Q^{(\chi)}_h
\left(1-i\gamma_5\tau_1\right) =
a\mu_\sigma + \tau_3 a\mu_\delta + N
- i\gamma_5\tau_1 \left(\mu_{\kappa h}+R\right) \ .
\ee

 For the SU(3) Yang-Mills gauge field we apply the {\em tree-level
 improved Symanzik} (tlSym) action which belongs to a one-parameter
 family of actions obtained by renormalisation group considerations
 and in the Symanzik improvement scheme \cite{Symanzik}.
 Those actions also include, besides the usual $(1\times 1)$ Wilson loop
 plaquette term, planar rectangular $(1\times 2)$ Wilson loops:
\be\label{eq12}
S_g = \beta\sum_{x}\left(c_{0}\sum_{\mu<\nu;\,\mu,\nu=1}^4
\left\{1-\frac{1}{3}\,{\rm Re\,} U_{x\mu\nu}^{1\times 1}\right\}
+c_{1}\sum_{\mu\ne\nu;\,\mu,\nu=1}^4
\left\{1-\frac{1}{3}\,{\rm Re\,} U_{x\mu\nu}^{1\times 2}\right\}
\right) \ ,
\ee
 with the condition $c_{0}=1-8c_{1}$.
 For the tlSym action we have $c_1=-1/12$ \cite{WeiszWohlert}.

\subsection{Simulation algorithm}\label{sec2.2}

 For preparing the sequences of gauge configurations a {\em Polynomial
 Hybrid Monte Carlo} (PHMC) updating algorithm was used.
 This algorithm is based on multi-step (actually two-step) polynomial
 approximations of the inverse fermion matrix with stochastic
 correction in the update chain as described in
 Ref.~\cite{MontvayScholz}.
 It is based on the PHMC algorithm as introduced in
 Ref.~\cite{FrezzottiJansen}.
 The polynomial approximation scheme and the stochastic correction
 in the update chain is taken over from the two-step multi-boson
 algorithm of Ref.~\cite{Montvay:tsmb}.
 (For an alternative updating algorithm in QCD with $N_f=2+1+1$ quark
 flavours, which will be used for algorithmic comparisons in the future,
 see \cite{Chiarappa:211}.)

 For typical values of the approximation interval and polynomial orders
 on $16^3 \cdot 32$ lattices see Table \ref{tab1}.
 The notations are those of Ref.~\cite{MontvayScholz}: the approximation
 interval is $[\epsilon,\lambda]$, the orders of the polynomials
 $P_j\;(j=1,2)$ are $n_j$ and those of $\bar{P}_j\;(j=1,2)$ are
 $\bar{n}_j$, respectively.
 The simulations have been done with {\em determinant break-up} $n_B=2$.
 On the $12^3 \cdot 24$, for similar values of the pseudoscalar masses
 in lattice units, the orders $n_2$ and $\bar{n}_2$  are the same and
 the values of $n_1$ and $\bar{n}_1$ are somewhat smaller.

\section{Physical fields and currents}\label{sec3}

 The {\em physical quark fields}, which in the continuum limit are
 proportional to the renormalised quark fields of both flavours in
 the doublets, are obtained \cite{Frezzotti:tmqcd} by a chiral rotation
 from the fields in the lattice action in (\ref{eq01}) or from those
 defined in (\ref{eq03}) for the light doublet, and similarly in
 (\ref{eq08})-(\ref{eq10}) for the heavy doublet.
 On the $\chi$-basis we have
\bea\label{eq13}
\psi^{phys}_{l,x} &=& e^{\frac{i}{2}\omega_l\gamma_5\tau_3} \chi_{l,x} \ ,
\hspace*{3em}
\overline{\psi}^{\;phys}_{l,x} = \overline{\chi}_{l,x}
e^{\frac{i}{2}\omega_l\gamma_5\tau_3} \ ;
\\[0.5em]
\label{eq14}
\psi^{phys}_{h,x} &=& e^{\frac{i}{2}\omega_h\gamma_5\tau_1} \chi_{h,x} \ ,
\hspace*{3em}
\overline{\psi}^{\;phys}_{h,x} = \overline{\chi}_{h,x}
e^{\frac{i}{2}\omega_h\gamma_5\tau_1} \ .
\eea
 Since the transformations in (\ref{eq03}) and (\ref{eq10}) correspond
 to chiral rotations with $\omega_l=\frac{\pi}{2}$ and
 $\omega_h=\frac{\pi}{2}$, respectively, we have with
\be\label{eq15}
\bar{\omega}_l \equiv \omega_l - \frac{\pi}{2} \ , \hspace*{3em}
\bar{\omega}_h \equiv \omega_h - \frac{\pi}{2}
\ee
 the relations
\bea\label{eq16}
\psi^{phys}_{l,x} &=& e^{\frac{i}{2}\bar{\omega}_l\gamma_5\tau_3} 
\psi_{l,x} \ , \hspace*{3em}
\overline{\psi}^{\;phys}_{l,x} = \overline{\psi}_{l,x}
e^{\frac{i}{2}\bar{\omega}_l\gamma_5\tau_3} \ ;
\\[0.5em]
\label{eq17}
\psi^{phys}_{h,x} &=& e^{\frac{i}{2}\bar{\omega}_h\gamma_5\tau_1} 
\psi_{h,x} \ , \hspace*{3em}
\overline{\psi}^{\;phys}_{h,x} = \overline{\psi}_{h,x}
e^{\frac{i}{2}\bar{\omega}_h\gamma_5\tau_1} \ .
\eea
 Since the simulations are usually performed near {\em full twist}
 corresponding to $\omega_l=\omega_h=\frac{\pi}{2}$, the modified
 twist angles are close to zero:
\be\label{eq18}
\bar{\omega}_l \simeq 0 , \ \hspace*{3em}
\bar{\omega}_h \simeq 0 \ . 
\ee
 Therefore, near full twist the $\psi$-fields are approximately equal
 to the physical quark fields.
 At full twist the use of the $\psi$-basis is advantageous because
 the formulas are simpler than in the $\chi$-basis.

 The definition of the twist angles is not unique.
 There are different viable possibilities to define them and the
 {\em critical hopping parameters} corresponding to them (see, for
 instance, \cite{Farchioni:twistangle}, \cite{Farchioni:dbw2},
 \cite{AokiBaer}, \cite{Bietenholz:overlapversustm},
 \cite{SharpeWu:nlochpt}, \cite{Frezzotti:MPR}, \cite{Sharpe:kappacr}).

 Here, for the light doublet, we use the definition based on the
 requirement of parity conservation for some matrix element of the
 physical vector and axialvector current, as first introduced in
 \cite{Farchioni:twistangle}, \cite{Farchioni:dbw2} and numerically
 studied in detail in \cite{Farchioni:dbw2two}.
 For this let us introduce the bare vector- and axialvector bilinears
\be\label{eq19}
V^a_{l,x\mu} \equiv \overline{\chi}_{l,x} \half\tau_a\gamma_\mu
\chi_{l,x} \ , \hspace{3em}
A^a_{l,x\mu} \equiv \overline{\chi}_{l,x} \half\tau_a\gamma_\mu\gamma_5
\chi_{l,x}
\hspace{3em}
(a=1,2) \ .
\ee
 The twist angle is introduced as the chiral rotation angle between the
 renormalised (physical) chiral currents:
\bea\label{eq20}
 \hat{V}^a_{l,x\mu} &=& Z_{lV} V^a_{l,x\mu}\,\cos\omega_l\, + 
\epsilon_{ab} \, Z_{lA} A^b_{l,x\mu}\,\sin\omega_l \ ,
\\[0.5em]\label{eq21}
 \hat{A}^a_{l,x\mu} &=& Z_{lA} A^a_{l,x\mu}\,\cos\omega_l\, +
\epsilon_{ab} \, Z_{lV} V^b_{l,x\mu}\,\sin\omega_l\,
\eea
 where only charged currents are considered ($a$=1,\,2), $\epsilon_{ab}$
 is the antisymmetric unit tensor and $Z_{lV}$ and $Z_{lA}$ are the
 multiplicative renormalisation factors of the vector and axialvector
 current, respectively.
 The exact requirements defining $\omega_l$ (and also yielding the
 value of $Z_{lA}/Z_{lV}$) is taken to be
\be\label{eq22}
\langle\, 0\,|\, \hat{V}^+_{l,x,\mu=0}\,|\,\pi^-\,\rangle\;=\; 0\ ,
\hspace{3em}
\langle\, 0\,|\, \hat{A}^+_{l,x,\mu=1,2,3}\,|\,\rho^-\,\rangle\;=\; 0\ .
\ee

 For the heavy doublet, in principle, one could translate and use this
 construction, too, but for applications in the kaon and D-meson sector
 it is more natural to consider bilinears between the light and the
 heavy doublet.
 In addition, inside the heavy doublet, due to the off-diagonal twist,
 one also would have to consider disconnected quark contributions which
 are absent in the light-heavy sector.
 Let us introduce the bare vector-, axialvector-, scalar- and
 pseudoscalar bilinears in the $K^+$- and $D^0$-sector as
\bea\label{eq23}
V_{K^+,x\mu} & \equiv & \overline{\chi}_{s,x} \gamma_\mu 
\chi_{u,x} \ , \hspace{3em}
A_{K^+,x\mu} \equiv \overline{\chi}_{s,x} \gamma_\mu \gamma_5 
\chi_{u,x} \ ,
\\[0.5em]\label{eq24}
S_{K^+,x} & \equiv & \overline{\chi}_{s,x} \chi_{u,x} \ , 
\hspace{4.5em}
P_{K^+,x} \equiv \overline{\chi}_{s,x} \gamma_5 \chi_{u,x} \ , 
\\[0.5em]\label{eq25}
V_{D^0,x\mu} & \equiv & \overline{\chi}_{c,x} \gamma_\mu 
\chi_{u,x} \ ,  \hspace{3em}
A_{D^0,x\mu} \equiv \overline{\chi}_{c,x} \gamma_\mu \gamma_5 
\chi_{u,x} \ ,
\\[0.5em]\label{eq26}
S_{D^0,x} & \equiv & \overline{\chi}_{c,x} \chi_{u,x} \ , 
\hspace{4.5em}
P_{D^0,x} \equiv \overline{\chi}_{c,x} \gamma_5 \chi_{u,x} \ ,
\eea
 and similarly for the $K^0$- and $D^-$-sector by changing $u \to d$.
 Denoting the kaon- and D-meson-doublet by $K \equiv (K^+\;\; K^0)$
 and $D \equiv (D^0\;\; D^-)$, respectively,
 and introducing $\breve{K} \equiv (K^+\; -\hspace*{-0.2em}K^0)$ and
 $\breve{D} \equiv (D^0\; -\hspace*{-0.2em}D^-)$, the renormalised
 vector and axialvector currents of the kaon doublet are given by
\bea\label{eq27}
\hat{V}_{K,x\mu} = & & 
\cos\frac{\omega_l}{2}\cos\frac{\omega_h}{2} Z_V V_{K,x\mu} +
\sin\frac{\omega_l}{2}\sin\frac{\omega_h}{2} Z_V V_{\breve{D},x\mu}
\nonumber\\[0.5em] & & +\;
i\sin\frac{\omega_l}{2}\cos\frac{\omega_h}{2} Z_A A_{\breve{K},x\mu} -
i\cos\frac{\omega_l}{2}\sin\frac{\omega_h}{2} Z_A A_{D,x\mu} \ ,
\\[0.5em]\label{eq28}
\hat{A}_{K,x\mu} = & & 
\cos\frac{\omega_l}{2}\cos\frac{\omega_h}{2} Z_A A_{K,x\mu} +
\sin\frac{\omega_l}{2}\sin\frac{\omega_h}{2} Z_A A_{\breve{D},x\mu}
\nonumber\\[0.5em] & & +\;
i\sin\frac{\omega_l}{2}\cos\frac{\omega_h}{2} Z_V V_{\breve{K},x\mu} -
i\cos\frac{\omega_l}{2}\sin\frac{\omega_h}{2} Z_V V_{D,x\mu} \ .
\eea
 Analogously for the scalar bilinears:
\bea\label{eq29}
\hat{S}_{K,x} = & & 
\cos\frac{\omega_l}{2}\cos\frac{\omega_h}{2} Z_S S_{K,x} -
\sin\frac{\omega_l}{2}\sin\frac{\omega_h}{2} Z_S S_{\breve{D},x}
\nonumber\\[0.5em] & & +\;
i\sin\frac{\omega_l}{2}\cos\frac{\omega_h}{2} Z_P P_{\breve{K},x} +
i\cos\frac{\omega_l}{2}\sin\frac{\omega_h}{2} Z_P P_{D,x} \ ,
\\[0.5em]
\label{eq30}
\hat{P}_{K,x} = & & 
\cos\frac{\omega_l}{2}\cos\frac{\omega_h}{2} Z_P P_{K,x} -
\sin\frac{\omega_l}{2}\sin\frac{\omega_h}{2} Z_P P_{\breve{D},x}
\nonumber\\[0.5em] & & +\;
i\sin\frac{\omega_l}{2}\cos\frac{\omega_h}{2} Z_S S_{\breve{K},x} +
i\cos\frac{\omega_l}{2}\sin\frac{\omega_h}{2} Z_S S_{D,x} \ .
\eea
 Similar relations hold in the D-meson doublet, too.
 (\ref{eq27})-(\ref{eq28}) show that near full twist
 $\omega_{l,h} \simeq \pi/2$ all four terms on the right hand sides
 have roughly equal coefficients.

 The requirement of parity symmetry in the isotriplets (pions,
 rho-mesons) allows to fix the twist angle $\omega_l$, cf. (\ref{eq22}).
 In the case of the heavy-light isodoublet one has to take into account
 the mixing between the kaons and D-mesons.
 In this case the twist angle $\omega_h$ (and $\omega_l$) can be fixed
 by requiring conservation of parity and/or flavour symmetry. 

 The equations in (\ref{eq22}) follow by considering
 \cite{Farchioni:twistangle}, \cite{Farchioni:dbw2},
 \cite{Farchioni:dbw2two} the (vanishing) vector-current-pseudoscalar
 and axialvector-current-vector-current correlators, which turns out to
 be the most convenient choice for fixing the twist angle in the light
 sector.
 In the heavy-light sector the mixing patterns for currents and scalar
 bilinears are similar, so any combination of operators gives similar
 expressions.
 However, correlators only made up of scalar bilinears are expected to
 give a better signal, so we concentrate on this case for the
 discussion. 
 Considering the upper components, four bilinears $P_{K^+}$, $P_{D^0}$,
 $S_{K^+}$, $S_{D^0}$ and the respective charge-conjugated versions
 must be included in the analysis.
 We define a four-dimensional vector of the multiplicatively
 renormalised bilinears
\be\label{eq31}
{\cal V}=\left ( \!\!\! \begin{array}{c} 
Z_P P_{K^+} \\ Z_P P_{D^0} \\ Z_S S_{K^+} \\ Z_S S_{D^0} \end{array}
\!\!\! \right ) \quad\quad
\bar {\cal V}=\left (-Z_P P_{K^-}, -Z_P P_{\bar D^0}, Z_S S_{K^-}, 
Z_S S_{\bar D^0} \right  )
\ee
 and analogously the vector $\hat{\cal V}$ of the {\em fully}
 renormalised bilinears according to (\ref{eq29}), (\ref{eq30})
 (and the analogous equations for the $D$-mesons).
 (\ref{eq29}) and (\ref{eq30}) can be then reformulated in a
 compact notation as
\be\label{eq32} 
\hat{\cal V}={\cal M}\,{\cal V}\ ,
\quad\quad\bar{\hat{\cal V}} = \bar{\cal{V}} \,{\cal M}^{-1}
\ee
 with the $4\times4$ matrix ${\cal M}$ given by
\be\label{eq33}
{\cal M}(\omega_l,\omega_h)\:=\:\left (\begin{array}{cccc}
\phantom{-}\cos\alphah\cos\betah & -\sin\alphah\sin\betah &
 i\sin\alphah\cos\betah & i\sin\betah\cos\alphah \\[1mm]
 -\sin\alphah\sin\betah & \phantom{-}\cos\alphah\cos\betah &
 i\sin\betah\cos\alphah & i\sin\alphah\cos\betah \\[1mm]
i\sin\alphah\cos\betah & i\sin\betah\cos\alphah &
 \phantom{-}\cos\alphah\cos\betah & -\sin\alphah\sin\betah \\ [1mm]
i\sin\betah\cos\alphah & i\sin\alphah\cos\betah &
 -\sin\alphah\sin\betah & \phantom{-}\cos\alphah\cos\betah
\end{array}\right )\ .
\ee
 ${\cal M}$ is the unitary matrix describing the mixing pattern between
 the kaon and D-meson doublets.
 One can easily see that
\be\label{eq34}
{\cal M}^T={\cal M} \quad \mbox{and}\
{\cal M}^\dagger(\omega_l,\omega_h)={\cal M^*}(\omega_l,\omega_h)
={\cal M}(-\omega_l,-\omega_h)={\cal M}^{-1}(\omega_l,\omega_h) \ .
\ee
 (The last equality is expected since reversing the sign of the angles
 corresponds to the inverse chiral transformation).
 One can at this point define a {\em correlator matrix} in the
 kaon-D-meson sector by
\be\label{eq35}
{\cal C}\:=\:\langle {\cal V}\otimes \bar {\cal V}\rangle
\ee
 (for example, ${\cal C}_{11}\equiv -Z_P^2\langle P_{K^+} P_{K^-}\rangle$)
 and its fully renormalised version
 $\hat {\cal C}=\langle \hat{\cal V}\otimes \bar {\hat {\cal V}}\rangle$.
 One has
\bea\label{eq36}
\hat {\cal C}&=&{\cal M}(\omega_l,\omega_h)\, {\cal C}\,
{\cal M}^{-1}(\omega_l,\omega_h)\ ,\\
{\cal C}&=&{\cal M}^{-1}(\omega_l,\omega_h) \, \hat{\cal C}\, 
{\cal M}(\omega_l,\omega_h) \ .
\label{eq37}
\eea
 Restoration of parity- and flavour-symmetry implies that
 $\hat {\cal C}$ is a diagonal matrix with ${\cal M}(\omega_l,\omega_h)$
 the matrix realizing the diagonalisation.
 The off-diagonal elements of the matrix equation (\ref{eq36}) 
 can be in principle used to determine the angles $\omega_l$ and
 $\omega_h$, while the diagonal elements give the physical correlators
 from which e.g. the masses can be obtained.
 Of course, in general, parity and flavour can only be restored up to
 ${\cal O}(a)$ violations.

 Taking also into account the residual discrete symmetries possessed by
 the action defined by (\ref{eq01}) and (\ref{eq08})-(\ref{eq09}),
 the only non-trivial conditions are obtained by imposing
 the vanishing of the flavour violating matrix elements 
 $\hat {\cal C}_{12}$, $\hat {\cal C}_{34}$ and transposed. 
 Defining for brevity $s_l=\sin\alphah$, $s_h=\sin\betah$,
 $c_l=\cos\alphah$, $c_h=\cos\betah$, the conditions are:
\bea
\label{eq38}
& & \hat{\cal C}_{12}+ \hat {\cal C}_{21}\:=\:
\left[(\cli\cprime)^2+(\sli\sprime)^2\right]
 ({\cal C}_{12}+{\cal C}_{21})
+ \left[(\sli\cprime)^2+(\sprime \cli)^2\right]
 ({\cal C}_{34}+{\cal C}_{43})
\nonumber \\[1mm]
& & -2\cli\cprime \sli\sprime ({\cal C}_{11}
+{\cal C}_{22}-{\cal C}_{33}-{\cal C}_{44})
+i\sprime \cprime(\sli^2-\cli^2)({\cal C}_{13}
-{\cal C}_{31}+{\cal C}_{24}-{\cal C}_{42})
\nonumber \\[1mm]
& & +i\sli\cli({\sprime}^2-{\cprime}^2)({\cal C}_{23}
-{\cal C}_{32}+{\cal C}_{14}-{\cal C}_{41})\:=\: 0\ , \\[2mm]
\label{eq39}
& & \hat {\cal {\cal C}}_{34}+\hat {\cal {\cal C}}_{43}\:=\:
\left[(\cprime \sli)^2+(\cli \sprime)^2\right]
 ({\cal C}_{12}+{\cal C}_{21})
+\left[(\cli\cprime)^2+(\sli\sprime)^2\right]
 ({\cal C}_{34}+{\cal C}_{43})
\nonumber \\[1mm]
& & +2\cli\cprime \sli\sprime ({\cal C}_{11}
+{\cal C}_{22}-{\cal C}_{33}-{\cal C}_{44})
-i\sprime \cprime(\sli^2-\cli^2)({\cal C}_{13}
-{\cal C}_{31}+{\cal C}_{24}-{\cal C}_{42})
\nonumber \\[1mm]
& & -i\sli\cli({\sprime}^2-{\cprime}^2)({\cal C}_{23}
-{\cal C}_{32}+{\cal C}_{14}-{\cal C}_{41}) \:=\: 0 \ .
\eea
 The sum of the two above equations implies
\be\label{eq40}
{\cal C}_{12}+{\cal C}_{21}+{\cal C}_{34}+{\cal C}_{43}=0 \ .
\ee
 A non trivial relation for the renormalisation constants of the bilinears
 is obtained from~(\ref{eq40})
\be\label{eq41}
Z^2_P/Z^2_S=  \frac{\langle S_{K^+} S_{\bar D^0} \rangle 
+ \langle S_{D^0} S_{K^-} \rangle}
{\langle P_{K^+} P_{\bar D^0} \rangle 
+ \langle P_{D^0} P_{K^-} \rangle} \ ,
\ee
 which can be used for a non-perturbative determination of $Z_P/Z_S$.

 Using (\ref{eq40}), (\ref{eq38}) (or (\ref{eq39})) can be
 restated in a compact way as a relation between $\cot{\omega_h}$ and
 $\cot{\omega_l}$
\be\label{eq42}
 \cot{\omega_h} =
\frac{ {\cal C}_{11}+{\cal C}_{22}-{\cal C}_{33}-{\cal C}_{44}
      +i({\cal C}_{13}-{\cal C}_{31}+{\cal C}_{24}-{\cal C}_{42})\cot{\omega_l}}
     {({\cal C}_{12}+{\cal C}_{21}-{\cal C}_{34}-{\cal C}_{43})\cot{\omega_l}
      -i({\cal C}_{23}-{\cal C}_{32}+{\cal C}_{14}-{\cal C}_{41})}\ .
\ee
 This can be used to determine $\omega_h$ once $\omega_l$ is known.
 ($\omega_l$ can be obtained following the prescription of
 \cite{Farchioni:twistangle}, \cite{Farchioni:dbw2},
 \cite{Farchioni:dbw2two}).

 This discussion suggests that, especially near full twist where the
 mixing is maximal, the analysis of the masses in the kaon-D-meson
 sector should be performed by considering the 4-dimensional correlator
 matrix $\cal C$.

 For tuning the hopping parameters the {\em untwisted PCAC quark mass}
 is also very useful.
 In the light doublet it is defined by the PCAC-relation containing the
 axialvector current $A^a_{l,x\mu}$ in (\ref{eq19}) and the
 corresponding pseudoscalar density
 $P^a_{l,x}=\overline{\chi}_{l,x} \half\tau_a\gamma_5\chi_{l,x}$:
\be\label{eq43}
am_{\chi l}^{PCAC} \equiv
\frac{\langle \partial^\ast_\mu A^+_{l,x\mu}\, P^-_{l,y} \rangle}
{2\langle P^+_{l,x}\, P^-_{l,y} \rangle}
\ee
 where $\tau_\pm \equiv \tau_1 \pm i\tau_2$.
 The condition of full twist in the light quark sector obtained
 from~(\ref{eq22}) by setting $\omega_l=\pi/2$
 coincides~\cite{Farchioni:dbw2two} with $m_{\chi l}^{PCAC}=0$.

 In the heavy sector one can define an untwisted PCAC quark mass
 $m_{\chi h}^{PCAC}$, too.
 A natural definition is obtained by considering the axialvector Ward
 identity
\be\label{eq44}
\partial^\ast_\mu A^a_{h,x\mu} 
= 2am_{\chi h}^{PCAC}\,  P^a_{h,x} 
+ \left\{\begin{array}{cl} 2iZ_A^{-1}a\mu_\sigma S_{h,x}^0,&a=1 \\
                             0, & a=2 \\
                          (-2i)Z_A^{-1}a\mu_\delta P_{h,x}^0,&a=3
             \end{array}\right.
\ee
%
 where, in analogy with the light sector in (\ref{eq19}), we define
\be\label{eq45}
A^a_{h,x\mu} \equiv \overline{\chi}_{h,x} \half\tau_a\gamma_\mu\gamma_5\chi_{h,x}
\hspace{1em} (a=1,2,3) \ , \hspace{1em}
S^0_{h,x} \equiv \overline{\chi}_{h,x}\chi_{h,x} \ , \hspace{1em}
P^0_{h,x} \equiv \overline{\chi}_{h,x}\gamma_5\chi_{h,x}
\ .
\ee
 (Observe that for uniformity with the definition (\ref{eq43}) we
 incorporate a factor $Z_A^{-1}$ in the definition of the untwisted PCAC
 quark mass).
 The above identity could in principle be used to tune 
 $\omega_h$ to $\pi/2$ by imposing $am_{\chi h}^{PCAC}=0$.
 However, as already mentioned, the presence of disconnected
 contributions in the heavy sector are likely not to allow for precise
 determinations.

 One can consider also in this case the heavy-light sector.
 Here the axialvector Ward identities read
\bea\label{eq46}
\partial^\ast_\mu A_{K,x\mu} &=& 
(am_{\chi s}^{PCAC} + am_{\chi l}^{PCAC})\, P_{K,x\mu} +
 i Z_A^{-1}\, a\mu_l\, S_{\breve{K},x\mu} +  i Z_A^{-1}\, a\mu_\sigma\,
 S_{D,x\mu}\ \ \ \ \\
\partial^\ast_\mu A_{D,x\mu} &=& (am_{\chi c}^{PCAC} + 
am_{\chi l}^{PCAC})\, P_{D,x\mu} +
 i Z_A^{-1}\, a\mu_l\, S_{\breve{D},x\mu} +  i Z_A^{-1}\, a\mu_\sigma\,
 S_{K,x\mu}\ .\ \ \ \
\label{eq47}
\eea
 The solution of the over-determined linear system, obtained by taking
 a suitable matrix element (for instance,
 $\langle \partial^\ast_\mu A_{K^+,x\mu}\, P_{K^-,y} \rangle$),
 allows to determine numerically (together with (\ref{eq43}))
 the untwisted PCAC mass of the heavy quarks  $m_{\chi c}^{PCAC}$,
 $m_{\chi s}^{PCAC}$ and the renormalisation factor $Z_A$. 
 The condition of full twist in the heavy doublet can be written as
\be\label{eq48}
 m_{\chi h}^{PCAC}\equiv m_{\chi c}^{PCAC}+ m_{\chi s}^{PCAC} = 0 \ .
\ee

 The quark masses defined by (\ref{eq43}) and (\ref{eq46})-(\ref{eq47})
 are untwisted components of bare quark masses.
 The physical quark masses can be obtained by the corresponding
 PCAC-relations of the renormalised currents and densities:
\bea\label{eq49}
& & am_l^{PCAC} \equiv
\frac{\langle \partial^\ast_\mu \hat{A}^+_{l,x\mu}\, \hat{P}^-_{l,y} \rangle}
{2\langle \hat{P}^+_{l,x}\, \hat{P}^-_{l,y} \rangle} \ ,
\\[0.5em]\label{eq50}
& & am_s^{PCAC} + am_l^{PCAC} \equiv
\frac{\langle \partial^\ast_\mu \hat{A}_{K^+,x\mu}\, \hat{P}_{K^-,y} \rangle}
{\langle \hat{P}_{K^+,x}\, \hat{P}_{K^-,y} \rangle} \ ,
\\[0.5em]\label{eq51}
& & am_c^{PCAC} + am_l^{PCAC} \equiv
\frac{\langle \partial^\ast_\mu \hat{A}_{D^+,x\mu}\, \hat{P}_{D^-,y} \rangle}
{\langle \hat{P}_{D^+,x}\, \hat{P}_{D^-,y} \rangle} \ .
\eea
 They are related to the bare quark masses by
\bea\label{eq52}
m_l^{PCAC}&=&Z_P^{-1}\,\sqrt{(Z_A\,m_{\chi l}^{PCAC})^2+\mu_l^2} \ ,
\\[1mm]
m_{c,s}^{PCAC}&=&Z_P^{-1}\,\sqrt{(Z_A\,m_{\chi h}^{PCAC})^2+\mu_\sigma^2}\, 
\pm\,  Z_S^{-1}\,\mu_\delta \ .
\label{eq53}
\eea
%

\section{Numerical simulations}\label{sec4}

 Our main goal in this work is to gain experience with the tuning of
 lattice parameters for future large scale simulations.
 Based on our recent work with $N_f=2$ dynamical twisted mass Wilson
 fermion QCD simulations in Refs.~\cite{Farchioni:tmphases},
 \cite{Farchioni:dbw2}, \cite{Farchioni:latticespacing},
 \cite{Farchioni:dbw2two} and \cite{LargeScaleNf2}, the main emphasis
 is on the effects of the additional dynamical flavours $s$ and $c$.
 As in the $N_f=2$ case, we start with coarse lattices: lattice spacings
 about $a \simeq 0.2\,{\rm fm}$ on a $12^3 \cdot 24$ lattice and
 $a \simeq 0.15\,{\rm fm}$ on a $16^3 \cdot 32$ lattice.
 (This implies spatial lattice extensions of $L \simeq 2.4\,{\rm fm}$.)
 The parameters of our main runs are: on the $12^3 \cdot 24$ lattice
 $\beta=3.25,\; a\mu_l=0.01,\; a\mu_\sigma=0.315,\; a\mu_\delta=0.285$
 and on the $16^3 \cdot 32$ lattice $\beta=3.35,\; a\mu_l=0.0075,\;
 a\mu_\sigma=0.2363,\; a\mu_\delta=0.2138$.
 The statistics is between 500 and 1100 PHMC trajectories of length
 $0.4$.
 (Of course, in order to find the appropriate parameters, we also had to
 perform at the beginning several additional short runs which we do not
 include here.)

 The tuning to full twist of the theory with an additional heavy doublet
 is complicated by the fact that {\em two} independent parameters
 $\kappa_{l}$ and $\kappa_{h}$ must be set to their respective critical
 values, using e.g. for the heavy sector the procedure outlined in the
 previous section. 
 However, it can be shown that in the continuum limit the deviation of
 the two critical hopping parameters $\kappa_{l,cr}$ and $\kappa_{h,cr}$
 goes to zero as ${\cal O}(a)$.
 An argument is given in the Appendix.
 This suggests to tune $\kappa_{l}$ to the value where
 $m^{PCAC}_{\chi l}=0$ with $\kappa_{h}=\kappa_{l}$: in this situation
 $m_{\chi h}^{PCAC}={\cal O}(a)$. 
 Observe that since the average quark mass in the heavy sector is
 typically large, the ${\cal O}(a)$ error is expected not to affect the
 full twist improvement in the sense of~\cite{FREZZOTTI-ROSSI},
 while it is critical to have good tuning in the light quark sector.
 This can be checked by computing $\omega_h$ as suggested in the
 previous section and verifying $\omega_h\approx \pi/2$.

 In view of this, we have set the two hopping parameters to be equal in
 our main runs:
 $\kappa\equiv\kappa_l=\kappa_h$.
 (In a few additional runs we checked that small individual changes of
 $\kappa_h$ by $\Delta\kappa_h \simeq 0.001$ do not alter any of the
 qualitative conclusions.)

\begin{figure}
\begin{center}
{\Large\bf Figures}
\end{center}
\vspace*{1em}
\begin{center}
\vspace*{0.10\vsize}
\begin{minipage}[c]{1.0\linewidth}
\hspace{0.03\hsize}
\includegraphics[angle=-90,width=0.90\hsize]
 {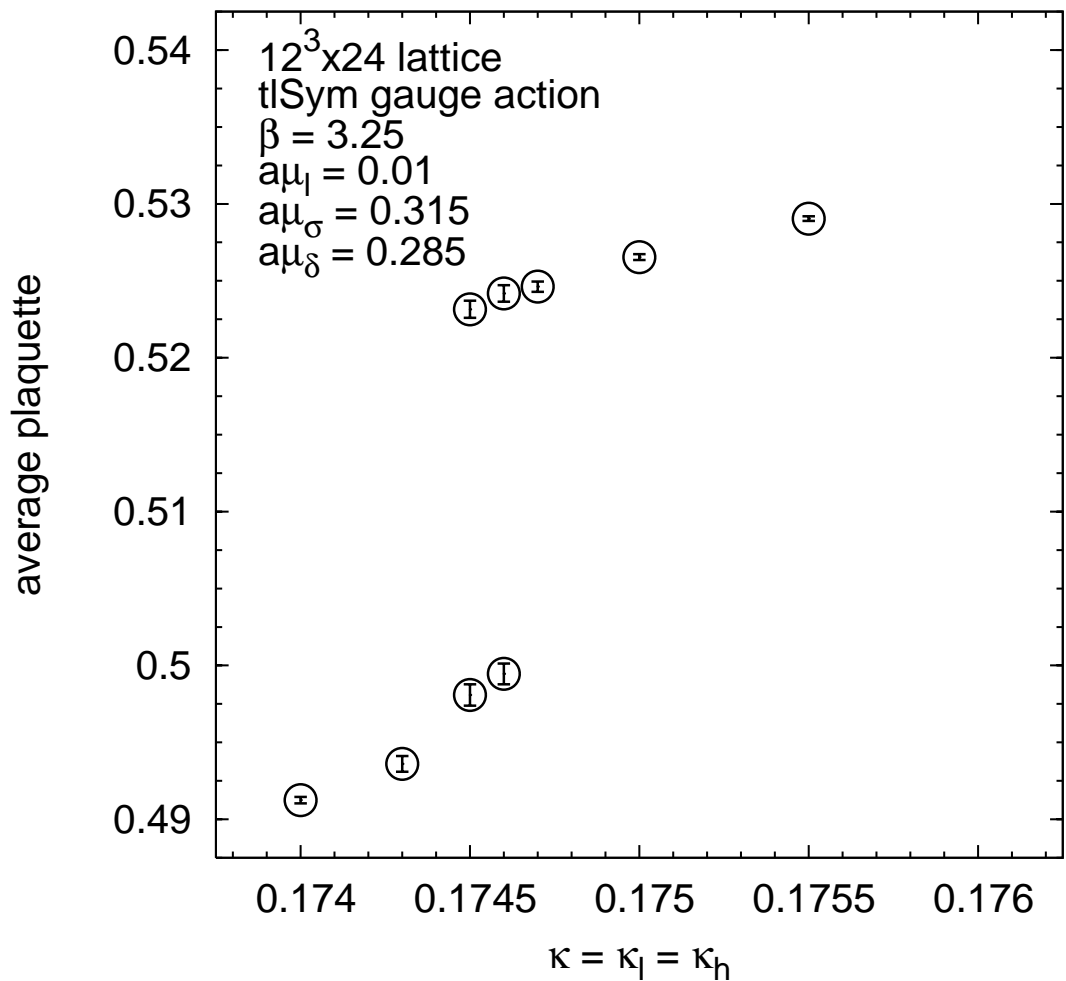}
\end{minipage}
\end{center}
\vspace*{-2em}
\begin{center}
\parbox{0.8\linewidth}{\caption{\label{fig_b325_plaq}\em
 The average plaquette on $12^3 \cdot 24$ lattice at
 $\beta=3.25,\;a\mu_l=0.01,\;a\mu_\sigma=0.315,\;a\mu_\delta=0.285$
 as a function of $\kappa\equiv\kappa_l=\kappa_h$.
 }}
\end{center}
\end{figure}

\begin{figure}
\begin{center}
\vspace*{0.10\vsize}
\begin{minipage}[c]{1.0\linewidth}
\hspace{0.03\hsize}
\includegraphics[angle=-90,width=0.90\hsize]
 {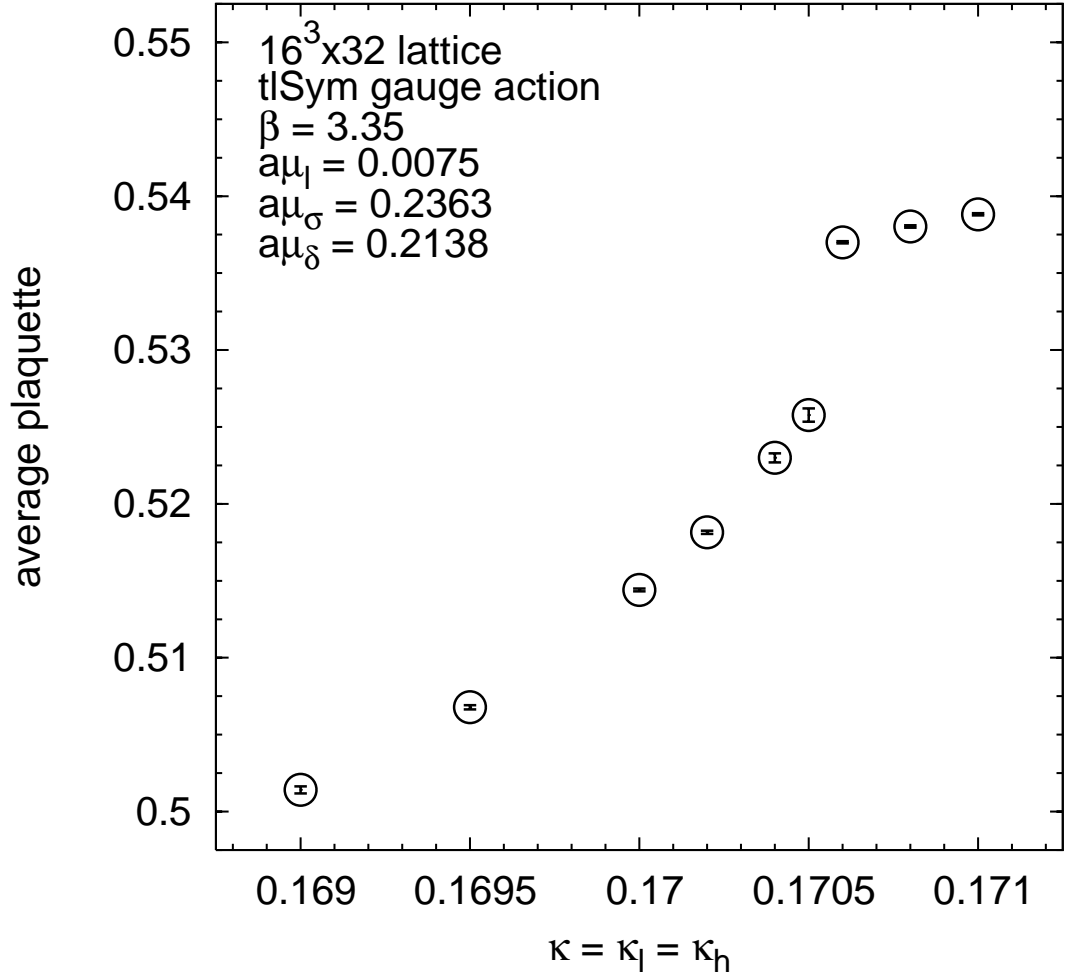}
\end{minipage}
\end{center}
\vspace*{-2em}
\begin{center}
\parbox{0.8\linewidth}{\caption{\label{fig_b335_plaq}\em
 The average plaquette on $16^3 \cdot 32$ lattice at
 $\beta=3.35,\;a\mu_l=0.0075,\;a\mu_\sigma=0.2363,\;a\mu_\delta=0.2138$
 as a function of $\kappa\equiv\kappa_l=\kappa_h$.
 }}
\end{center}
\end{figure}

 The average plaquette values as a function of the hopping parameter
 $\kappa_l=\kappa_h=\kappa$, for fixed values of the twisted masses, are
 shown by Figures \ref{fig_b325_plaq} and \ref{fig_b335_plaq} on the
 $12^3 \cdot 24$ and $16^3 \cdot 32$ lattices, respectively.
 On the $12^3 \cdot 24$ lattice a strong metastability is observed
 for $0.1745 \leq \kappa \leq 0.1747$, which we interpret as the
 manifestation of a {\em first order phase transition}.
 This behaviour agrees with one of the scenarios predicted by ChPT
 including leading lattice artifacts \cite{SharpeSingleton},
 \cite{Munster:phases}, \cite{Scorzato:phases}, \cite{SharpeWu:phases}.
 It has also been observed in our previous simulations, for instance, in
 \cite{Farchioni:tmphases}.
 On the $16^3 \cdot 32$ lattice no metastability could be observed,
 although there is a sharp rise of the average plaquette value
 between $\kappa=0.1705$ and $\kappa=0.1706$.
 This may also signal a (weaker) first order phase transition or
 a {\em cross-over}.
 To decide among these two possibilities, in principle, an
 investigation of the infinite volume behaviour would be necessary.
 In practice, in a finite volume, the effects of a real first order
 phase transition and a cross-over are similar.

\begin{figure}
\begin{center}
\vspace*{0.10\vsize}
\begin{minipage}[c]{1.0\linewidth}
\hspace{0.03\hsize}
\includegraphics[angle=-90,width=0.90\hsize]
 {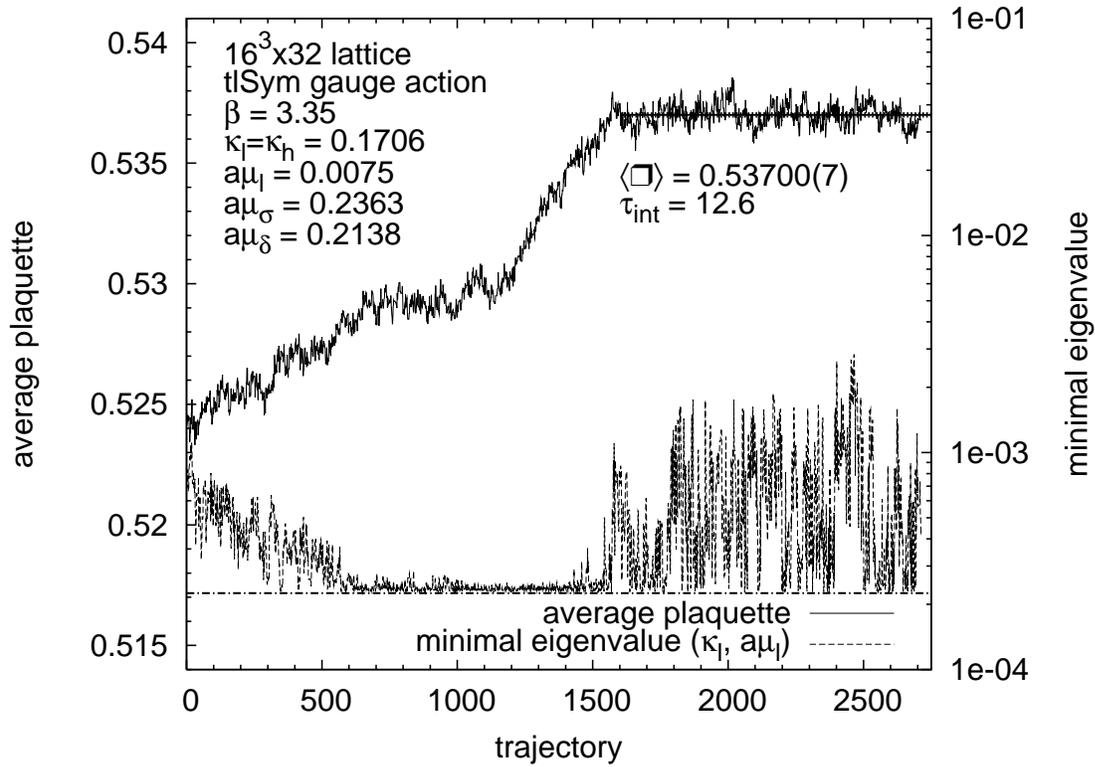}
\end{minipage}
\end{center}
\vspace*{-2em}
\begin{center}
\parbox{0.8\linewidth}{\caption{\label{fig_b335_k1706}\em
 Run history on a $16^3 \cdot 32$ lattice at
 $\beta=3.35,\;a\mu_l=0.0075,\;a\mu_\sigma=0.2363,\;a\mu_\delta=0.2138,\;
 \kappa_l=\kappa_h=0.1706$.
 This run started from a previous one at $\kappa=0.1705$.
 On the horizontal axis the number of PHMC-trajectories (of length
 $\Delta\tau=0.4$) is given.
 The average plaquette (upper curve, left scale) and the smallest eigenvalue
 of the squared preconditioned fermion matrix $\lambda_{min}$ (lower curve,
 right scale) are shown.
 The horizontal lines indicate the average plaquette after equilibration
 and the absolute minimum of $\lambda_{min}$, respectively.
 }}
\end{center}
\end{figure}

 We emphasize that this observed behaviour is not related to some
 imperfection of the simulation algorithm.
 Due to the positive twisted masses the eigenvalues of the fermion
 matrix have a positive lower bound.
 Therefore, we could choose the HMC step size small enough in order
 that the molecular dynamical force does not become too large.
 The behaviour of the system when crossing the phase transition region
 is nicely illustrated by the run history in Figure
 \ref{fig_b335_k1706}.
 One can recognize three stages in the plot: metastable start at  
 $m_{\chi l}^{PCAC} > 0$; crossing; stable thermalization at
 $m_{\chi l}^{PCAC} < 0$.
 A high concentration of small eigenvalues occurs during the crossing,
 because a large portion of the Dirac spectrum (actually all the  
 physically relevant eigenvalues) is moving from the right half complex
 plane with ${\rm Re}\lambda>0$ to the left one with
 ${\rm Re}\lambda<0$.

 We determined several quantities in both the pion- and kaon-sector.
 The values of some of them are collected in Tables \ref{tab2} and
 \ref{tab3}.
 As in our previous work, we determined the lattice spacing from the
 quark force by the Sommer scale parameter \cite{Sommer:scale}
 assuming $r_0 \equiv 0.5\,{\rm fm}$.
 Taking the values for positive untwisted PCAC quark masses
 ($am_{\chi l}^{PCAC} > 0$), we get for $\beta=3.25$ on the
 $12^3 \cdot 24$ lattice $a(\beta=3.25) \simeq 0.20\,{\rm fm}$ and for
 $\beta=3.35$ on the $16^3 \cdot 32$ lattice
 $a(\beta=3.35) \simeq 0.15\,{\rm fm}$.
 These correspond to $a^{-1} \simeq 1.0\,{\rm GeV}$ and
 $a^{-1} \simeq 1.3\,{\rm GeV}$, respectively.

 It follows from the data in Tables \ref{tab2} and \ref{tab3} that the
 pion, and hence the $u$- and $d$-quark masses, are not particularly
 small in our runs.
 Considering only the points with positive untwisted PCAC quark mass
 ($am_{\chi l}^{PCAC} > 0$) outside the metastability region at
 $\beta=3.25$ we have $m_\pi \geq 670\,{\rm MeV}$.
 At $\beta=3.35$ the corresponding value is
 $m_\pi \geq 450\,{\rm MeV}$.
 (The points with $am_{\chi l}^{PCAC} < 0$ have $m_\pi \geq 530\,{\rm MeV}$
 and $m_\pi \geq 560\,{\rm MeV}$ for the two $\beta$-values,
 respectively, but they are usually not considered for large scale
 simulations because of the strongly fluctuating small eigenvalues as
 shown, for instance, by Fig.~\ref{fig_b335_k1706}.)

 The kaon masses are also given Tables \ref{tab2} and \ref{tab3}.
 Let us note that, in the Frezzotti-Rossi formulation of the split-mass
 doublet we use, the masses in the kaon doublet (and D-meson doublet)
 are exactly degenerate.
 This follows from an exact symmetry of the lattice action defined in
 Section~\ref{sec2.1} (both in the $\chi$- and $\psi$-basis of quark
 fields) namely, simultaneous multiplication by an isospin matrix and
 space reflection:
\be\label{eq54}
S:\left\{
\begin{array}{l}
\mbox{light:} \hspace*{1em} \mbox{Parity}\otimes\tau_1:\left\{
\begin{array}{l}
u(x)\rightarrow  \phantom{-}\gamma_0\,d(x_P) \\
d(x)\rightarrow  \phantom{-}\gamma_0\,u(x_P) \\
\end{array} \right .
\\[7mm]
\mbox{heavy:  Parity}\otimes\tau_3:\left\{
\begin{array}{l}
c(x)\rightarrow  \phantom{-}\gamma_0\,c(x_P) \\
s(x)\rightarrow -\gamma_0\,s(x_P) \\
\end{array} \right .
\end{array}
\right .
\ee
 This {\em exact} symmetry exchanges the $u$-quark and the $d$-quark,
 hence the equality of the masses within kaon- and D-meson doublets
 follows.

 Let us note that in a recent publication \cite{Lewis:quenchedkaon}
 a non-zero kaon mass splitting has been calculated in the quenched
 approximation using another formulation \cite{PenaSintVladikas} of the
 mass-split doublet where both the twist and the mass splitting are
 in the same isospin direction.
 This formulation has, however, the disadvantage that the fermion
 determinant is non-real and therefore an unquenched computation is
 practically impossible at present.
 The difference in the presence and absence of the kaon mass splitting
 in the two formulations comes from the fact that the states with a
 given quark flavour correspond to different linear combinations here
 and there.

 Similarly to the pion masses, the kaon masses in Tables \ref{tab2} and
 \ref{tab3} are also higher than the physical value.
 In the points cited above for the pion mass we have: at
 $\beta=3.25$ and $\beta=3.35$ $m_K \geq 920\,{\rm MeV}$ and
 $m_K \geq 850\,{\rm MeV}$, respectively.
 The kaon mass can be easily lowered by tuning the mass parameters in
 the heavy doublet.
 In order to explore this we also performed simulations at
 $\beta=3.35,\;a\mu_l=0.0075$ on the $16^3 \cdot 32$ lattice with
 $a\mu_\sigma=a\mu_\delta=0.15$.
 For instance, at $\kappa_l=\kappa_h=0.17$ we got $am_\pi=0.4432(40)$
 and $am_K=0.5918(22)$.
 Comparing to the third line in Table \ref{tab3} one can see that both
 the pion and the kaon mass become smaller.
 In particular, the kaon mass is smaller by a factor of about $3/4$.
 This shows that the kaon mass can probably be tuned to its physical
 value if wanted.
 Another possibility is to do the chiral extrapolation by fixing,
 instead of $m_K$, the pion-kaon mass ratio $m_\pi/m_K$ to its physical
 value.

 The D-meson masses in Tables \ref{tab2} and \ref{tab3} are typically
 smaller than the physical value.
 In the points cited above for the pion and kaon masses we have: at
 $\beta=3.25$ and $\beta=3.35$ $m_D \simeq 1450\,{\rm MeV}$ and
 $m_D \simeq 1400\,{\rm MeV}$, respectively.
 $m_D$ can, in principle, also be tuned to its physical value.
 However, on coarse lattices the D-meson mass is close to the cut-off
 and, therefore, it is more reasonable to keep it smaller than the
 physical value in order to be well below the cut-off.
 In fact, in our runs the actual D-meson masses are already at the
 cut-off because we have $a^{-1} \simeq 1\,{\rm GeV}$ and
 $a^{-1} \simeq 1.3\,{\rm GeV}$ at $\beta=3.25$ and $\beta=3.35$,
 respectively.
 But on a fine lattice, say with $a^{-1} \simeq 4\,{\rm GeV}$, it
 will become possible to directly go to the physical value of $m_D$,
 too.

 The machinery for the twist angle in the heavy doublet $\omega_h$
 developed in Sec.~\ref{sec3} has been tested in a few runs, too.
 The formulas worked fine and the results turned out to be plausible.
 For instance, in the run at $\beta=3.35,\; \kappa_l=\kappa_h=0.1704,\;
 a\mu_l=0.0075,\; a\mu_\sigma=0.2363,\;a\mu_\delta=0.2138$ on a
 $16^3 \cdot 32$ lattice we obtained from 400 gauge configurations:
\bea
&& \omega_l/\pi =  0.0981(55), \hspace{2em}
\omega_h/\pi =  0.490(25) ,
\nonumber
\\[0.5em]\label{eq55}
&& Z_P/Z_S = 0.5739(65), \hspace{2em}
Z_A = 0.897(11), \hspace{2em}
Z_V = 0.5490(12) . \
\eea
 As one sees, $\omega_h$ is rather close to $\pi/2$ even if $\omega_l$
 is still far from it.
 This is a consequence of $\mu_\sigma \gg \mu_l$.
 Using the relation (valid in the continuum)
 $\cot(\omega_h)/\cot(\omega_l) = \mu_l/\mu_\sigma$ and the value of
 $\omega_l$ given above, one would get $\omega_h/\pi= 0.468$.
 The situation is very similar in the runs on a $12^3 \cdot 24$ lattice,
 too.
 For instance, in the run with largest untwisted mass of Table~\ref{tab2}
 at $\kappa_l=\kappa_h=0.1740_L$ we obtained:
\be\label{eq56}
\omega_l/\pi = 0.04298(34), \hspace{2em}
\omega_h/\pi = 0.4356(83), \hspace{2em}
Z_P/Z_S = 0.581(11) . \
\ee

 These results imply that putting the untwisted quark mass equal in the
 two sectors gives an elegant solution for tuning to full twist:
 one can just do the same as in the $N_f=2$ case.
 Due to the large twisted component in the heavy sector, the tuning of
 $\omega_h$ to $\pi/2$ is no problem at all: already at moderate values
 of $\omega_l$, $\omega_h$ is almost equal to $\pi/2$.

 Let us finally mention that using Chiral Perturbation Theory (ChPT)
 formulas one can also extrapolate from our simulation points to smaller
 pion- and kaon-masses.
 As a simple example, let us take the squared pion-kaon mass ratio in
 lowest order (LO) ChPT:
\be\label{eq57}
(m_\pi/m_K)^2 = \frac{2 m_{ud}}{m_{ud} + m_s} \ .
\ee
 In terms of our parameters we can set 
\be\label{eq58}
m_{ud} = \sqrt{(Z_A\, m_{\chi l}^{PCAC})^2 + \mu_l^2} \ , \hspace*{2em}
m_s    = \sqrt{(Z_A\, m_{\chi h}^{PCAC})^2 + (\mu_\sigma)^2}
- \frac{Z_P}{Z_S}\, \mu_\delta
\ee
 where $Z_P/Z_S$ is a fitted relative renormalisation factor.
 In our fits we set, for simplicity, $Z_A=0.897$ from (\ref{eq55}) and
 we also assumed
 $m_{\chi h}^{PCAC} = m_{\chi l}^{PCAC}$, which corresponds to the
 assumption $\kappa_{h,cr}=\kappa_{l,cr}$.
 The results for both $\beta$ values are shown in
 Figure~\ref{fig_mPi-mK_chpt}.
 Note that although these fits look rather good, clearly, the validity
 of Chiral Perturbation Theory in general has to be checked in further
 simulations at small values of $a$ and $m_\pi$.

 It turns out that the fitted values of $Z_P/Z_S$ are well below 1,
 namely $Z_P/Z_S \simeq 0.45$, which implies that, as also directly
 shown by our simulation data, the kaon mass reacts relatively weakly
 to the change of the bare quark mass difference parameter
 $a\mu_\delta$.
 The deviation of $Z_P/Z_S$ obtained in the LO-ChPT fit from
 the values in (\ref{eq55}) and (\ref{eq56}) might be due to lattice
 artifacts or/and to the fact that in (\ref{eq55})-(\ref{eq56}) no
 extrapolation to zero quark masses is performed.

 Note that the obtained values of $Z_P/Z_S$ do not satisfy the bound
 derived in \cite{FrezzottiRossi:split} which would ensure the
 positivity of the quark determinant, because in case of the
 $(c,s)$-doublet this bound is
 $Z_P/Z_S > (m_c - m_s)/(m_c + m_s) \simeq 0.85$.
 This means that there might be some gauge configuration where the
 determinant of the $(c,s)$-doublet is negative.
 However, such configurations have a very low probability and hence
 they practically never occur in Monte Carlo simulations.
 This is shown by the eigenvalues of the fermion matrix which never
 come close to zero: for the $(c,s)$-doublet in our
 simulations they always satisfy $\lambda_{min, h} > 0.01$.
 (This has to be compared to the minimal eigenvalues in the
 $(u,d)$-doublet which only satisfy $\lambda_{min, l} > 0.0001$.)

 It is remarkable that the minimum value of the interpolated curves in
 Figure \ref{fig_mPi-mK_chpt} are not far away from the physical value
 $(m_\pi/m_K)^2 \simeq 0.082$.
 This raises the interesting question, whether it would be possible to
 perform uncoventional chiral extrapolations from simulation data at 
 fixed twisted masses.

\section{Discussion}\label{sec5}

 The main conclusion of the present paper is that numerical simulations
 of QCD with unquenched $u$, $d$, $s$ and $c$ quarks are possible in the
 twisted-mass Wilson formulation.

 The PHMC updating algorithm with multi-step polynomial approximations
 and stochastic correction during the update turned out to be effective
 even in difficult situations near a first order phase transition (or
 cross-over).
 The autocorrelations of the quantities given in Tables \ref{tab2} and
 \ref{tab3} are typically of ${\cal O}(1)$ in number of
 PHMC-trajectories (most of the time of length 0.4), therefore, it is
 worth to analyse the gauge configurations after every trajectory.

 At $\beta=3.25$ (lattice spacing $a \simeq 0.20\,{\rm fm}$) on our
 $12^3 \cdot 24$ lattice we observed strong metastabilities suggesting
 a first order phase transition.
 This agrees with one of the scenarios predicted by ChPT including
 leading lattice artifacts \cite{SharpeSingleton}, \cite{Munster:phases},
 \cite{Scorzato:phases}, \cite{SharpeWu:phases} and has been observed
 previously in several QCD simulations with Wilson fermions
 \cite{Blum:transition}, \cite{JLQCD:transition}, \cite{Jansen:transition},
 \cite{Farchioni:tmphases}.
 At $\beta=3.35$ (lattice spacing $a \simeq 0.15\,{\rm fm}$) on our
 $16^3 \cdot 32$ lattice the phase transition becomes weaker but is
 still visible as a strong cross-over region with fast changes in
 several quantities.
 Compared to $N_f=2$ runs at similar lattice spacings the first order
 phase transition becomes stronger for $N_f=2+1+1$.
 (This agrees with the early observations in \cite{JLQCD:transition}.)

\begin{figure}
\begin{center}
\vspace*{0.10\vsize}
\begin{minipage}[c]{1.0\linewidth}
\hspace{0.03\hsize}
\includegraphics[angle=-90,width=0.85\hsize]
 {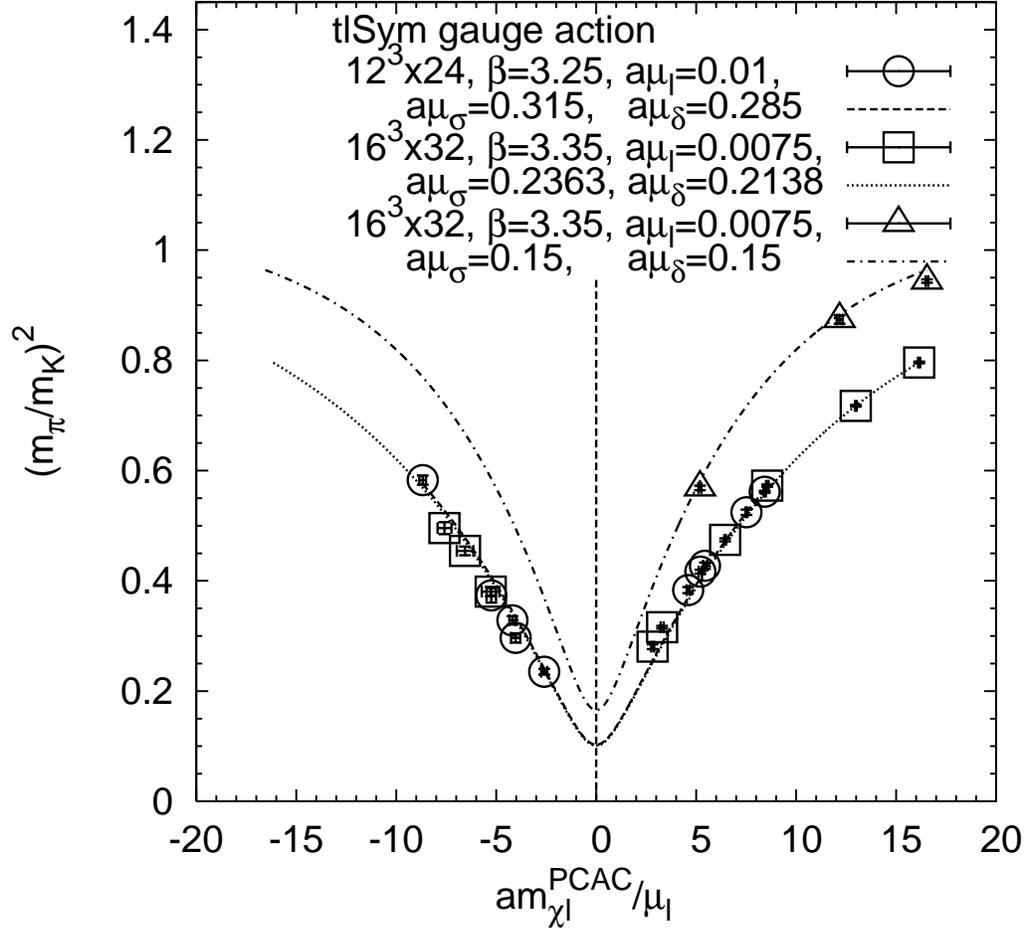}
\end{minipage}
\end{center}
\vspace*{-2em}
\begin{center}
\parbox{0.8\linewidth}{\caption{\label{fig_mPi-mK_chpt}\em
 Lowest order ChPT fit of the squared pion to kaon mass ratio as a
 function of the untwisted PCAC quark mass.
 Squares and triangles are data at $\beta=3.35,\;\mu_l=0.0075$ for
 $\mu_\sigma=0.2363,\;\mu_\delta=0.2138$ and
 $\mu_\sigma=0.15,\;\mu_\delta=0.15$, respectively.
 The fit to the formulas (\ref{eq57})-(\ref{eq58}) gives $Z_P/Z_S=0.446$.
 Circles are data at $\beta=3.25,\;\mu_l=0.01,\;\mu_\sigma=0.315,\;
 \mu_\delta=0.285$.
 The fit gives in this case $Z_P/Z_S=0.457$.
 }}
\end{center}
\end{figure}

 The smallest simulated pion mass in a stable point with positive
 untwisted PCAC quark mass $(am_{\chi l}^{PCAC} > 0)$ at $\beta=3.25$
 ($a \simeq 0.20\,{\rm fm}$) and $\beta=3.35$
 ($a \simeq 0.15\,{\rm fm}$) is
 $m_\pi \simeq 670\,{\rm MeV}$ and $m_\pi \simeq 450\,{\rm MeV}$,
 respectively.
 Our expectation based on the ChPT formulas and on our previous
 experience is that, for instance, on a $24^3 \cdot 48$ lattice with
 $a \simeq 0.10\,{\rm fm}$ the minimal pion mass at $a\mu=0.005$ will be
 somewhere in the range $270\,{\rm MeV} < m_\pi^{min} < 300\,{\rm MeV}$.
 This is because at vanishing twisted masses $m_\pi^{min}$ is going
 to zero as ${\cal O}(a)$ and for positive twisted mass the decrease
 is somewhat faster.
 (The lower value of the estimate corresponds to the minimum of the
 extrapolated curve in Figure \ref{fig_mPi-mK_chpt}.)

 The kaon mass in the present simulations is higher than the physical
 value but can probably be properly tuned by changing the twisted
 mass parameters in the $c$-$s$ doublet. 
 The D-meson mass is smaller than the physical value (i.e. the $c$-$s$
 mass splitting is smaller than in nature) but this is reasonable
 on coarse lattices in order to stay with it below the cut-off.
 On finer lattices (say, with $a \simeq 0.05\,{\rm fm}$) one can
 try to tune also the D-meson mass to its physical value.
 A possible difficulty in properly tuning the mass splittings in the
 $c$-$s$ doublet can be caused by the relative insensitivity of the
 masses to the bare mass-splitting parameter $a\mu_\delta$.
 This may imply the necessity of some extrapolations in the mass ratios.

 In case of the $c$-$s$-doublet the mass splitting is rather large
 because the renormalised quark masses satisfy
 $(m_c-m_s)/(m_c+m_s) \simeq 0.85$.
 Therefore it is important to take into account the mass splitting.
 For the $u$-$d$-doublet, well above the scale of $u$ and $d$ quark
 masses, the mass degeneracy can be considered as a good approximation,
 but even in this case we have in nature
 $(m_d-m_u)/(m_d+m_u) \simeq 0.28$.
 Hence also there, on a long run, the problem of the quark mass
 splitting within the doublet has to be tackled. 

 In summary, our experience in this paper is rather positive both for
 the twisted-mass Wilson fermion formulation and for the PHMC algorithm
 we are using.
 This opens the road for future large scale QCD simulations with
 dynamical $u$, $d$, $s$ and $c$ quarks.

\vspace*{1em}
\noindent
{\large\bf Acknowledgments}

\noindent
 We acknowledge helpful discussions with Roberto Frezzotti, Andrea
 Shindler and Urs Wenger.
 We thank the computer centers at DESY Hamburg and NIC at
 Forschungszentrum J{\"u}lich for providing us the necessary technical
 help and computer resources.
 This research has been supported by the DFG
 Sonderforschungsbereich/Transregio SFB/TR9-03 and in part by the
 EU Integrated Infrastructure Initiative Hadron Physics (I3HP) under
 contract RII3-CT-2004-506078 and also in part by the U.S. Department of
 Energy under contract number DE-AC02-98CH10886.
 The work of T.C. is supported by the DFG in the form of a
 Forschungsstipendium CH 398/1.

\newpage

\appendix
\noindent 
{\bf \Large Appendix}
\vspace*{1em}

\noindent 
 In the $N_f=2$ theory, one possible definition of the critical quark
 mass ${m_0}_{cr}(g_0,\mu)$ is given by the vanishing of the PCAC quark
 mass $m^{PCAC}_{\chi}$.
 Due to chirality breaking the latter gets shifted:
\be\label{eq59}
m^{PCAC}_\chi= m_0-a^{-1}f(g_0,am_0,a\mu)\ ,
\ee
 with $f$ a dimensionless function.
 On the basis of the symmetry of the action under parity $\times$
 $(\mu\rightarrow-\mu$) one can show that the additive renormalisation
 of the quark mass is {\em even} in $\mu$, and analyticity in turn
 implies
\be\label{eq60}
f(g_0,am_0,a\mu)= f(g_0,am_0)+O(\mu^2a^2)\ ,
\ee
 where $f(g_0,am_0)$ is the shift for ordinary $N_f=2$ QCD without
 twisted mass term.
 So the twisted mass term in the action only produces an ${\cal O}(a)$
 effect on the quark mass (with $g_0$ and $m_0$ held fixed):
\be\label{eq61}
m^{PCAC}_\chi=m_0-a^{-1}f(g_0,am_0)+{\cal O}(a)\ .
\ee

 The above argument can be easily generalized to the $N_f=2+1+1$ theory.
 Here one has to make a distinction between the two sectors:
\bea\label{eq62}
m^{PCAC}_{\chi l} &=& m_{0l}-a^{-1}
f_l(g_0,am_{0l},am_{0h},a\mu_l,a\mu_\sigma,a\mu_\delta) \ ,\\
m^{PCAC}_{\chi h} &=& m_{0h}-a^{-1}
f_h(g_0,am_{0h},am_{0l},a\mu_\sigma,a\mu_l,a\mu_\delta) \ .
\label{eq63}
\eea
 The functions $f_l$ and $f_h$ are in this case even in $\mu_{l}$,
 $\mu_{h}$  and $\mu_\delta$\footnote{An additional symmetry in the
 heavy sector is needed for the argument, namely
 $\chi_{h,x}\rightarrow\exp{\{i\frac{\pi}{2}\tau_1\}}\chi_{h,x}$,
 $\bar\chi_{h,x}\rightarrow\chi_{h,x} \exp{\{-i\frac{\pi}{2}\tau_1\}}$
 composed with $\mu_\delta\rightarrow-\mu_\delta$.}: similarly to
 $N_f=2$, the associated terms in the action only affect the additive
 renormalisation of the quark mass by ${\cal O}(a)$ terms.
 So we write:
\bea\label{eq64}
m^{PCAC}_{\chi l}&=&m_{0l}-a^{-1}f(g_0,am_{0l},am_{0h})+{\cal O}(a) \ ,\\
m^{PCAC}_{\chi h}&=&m_{0h}-a^{-1}f(g_0,am_{0h},am_{0l})+{\cal O}(a)\ ,
\label{eq65}
\eea
 where on the r.h.s. we have now the mass-shifts for the theory without
 twist and mass-splitting ($N_f=2+2$ QCD): here the distinction between
 the two sectors is immaterial.
 From eqs.~(\ref{eq64}), (\ref{eq65}) it
 follows immediately
\bea\label{eq66}
 m_{0l}=m_{0h}=m_0 \quad \Rightarrow \quad m^{PCAC}_{\chi h} =
m^{PCAC}_{\chi l}+{\cal O}(a)\ .
\eea
%



\clearpage
\begin{table}
\begin{center}
{\Large\bf Tables}
\end{center}
\vspace*{1em}
\begin{center}
\parbox{0.8\linewidth}{\caption{\label{tab1}\em
 Algorithmic parameters in two runs on a $16^3 \cdot 32$ lattice at
 $\beta=3.35,\;\kappa_l=\kappa_h=\kappa,\;a\mu_l=0.0075,\;
 a\mu_\sigma=0.2363,\;a\mu_\delta=0.2138$ and with determinant
 break-up $n_B=2$.
 The first line for a given $\kappa$ shows the pion mass and the
 parameters for the light doublet, the second line the kaon mass
 and the parameters for the heavy doublet.}}
\end{center}
\begin{center}
\renewcommand{\arraystretch}{1.2}
\begin{tabular}{*{8}{|r}|}
\hline
 \multicolumn{1}{|c|}{$\kappa$} &
 \multicolumn{1}{|c|}{$am_{\pi,K}$} &
 \multicolumn{1}{|c|}{$\epsilon$} &
 \multicolumn{1}{|c|}{$\lambda$} &
 \multicolumn{1}{|c|}{$n_1$} &
 \multicolumn{1}{|c|}{$\bar{n}_1$} &
 \multicolumn{1}{|c|}{$n_2$} &
 \multicolumn{1}{|c|}{$\bar{n}_2$}  \\
\hline\hline
 0.1690 & 0.8237(13) & 1.25e-2  & 25 & 70  & 110 & 120 & 160 \\
\hline
        & 0.9231(11) & 3.25e-2  & 26 & 50  & 80  & 90  & 130 \\
\hline\hline
 0.1705 & 0.3433(52) & 1.875e-4 & 25 & 220 & 320 & 800 & 930 \\
\hline
        & 0.6503(18) & 1.875e-2 & 26 & 60  & 100 & 120 & 160 \\
\hline
\end{tabular}
\end{center}
\end{table}

\clearpage
\begin{table}
\begin{center}
\parbox{0.8\linewidth}{\caption{\label{tab2}\em
 Selected results of the runs on a $12^3 \cdot 24$ lattice at
 $\beta=3.25,\; a\mu_l=0.01,\; a\mu_\sigma=0.315,\; a\mu_\delta=0.285$.
 The subscript on $\kappa=\kappa_l=\kappa_h$ denotes: $L$ for ``low''
 and $H$ for ``high'' plaquette phase, respectively.
}}
\end{center}
\begin{center}
\renewcommand{\arraystretch}{1.2}
\begin{tabular}{*{7}{|l}|}
  \hline
  $\kappa_l=\kappa_h$ & \,\,\,\,\,\,\, 
  $r_0/a$ & \,\,\,\, 
  $am_{\pi}$ &\,\,\,\, 
  $am_{\rho}$ & \,\,\,\, 
  $am_{K}$ & \,\,\,\, 
  $am_{D}$ & \,\,\,\,
  $am_{\chi l}^{PCAC}$
 \\ \hline
 $0.1740_L$ & 2.35(12)  & 0.7110(21) & 0.9029(27) & 0.9487(16) & 1.4858(75) &
 0.08432(56)
 \\ \hline
 $0.1743_L$ & 2.279(56) & 0.6718(59) & 0.8756(30) & 0.9277(22) & 1.4543(99) &
 0.07515(45)
 \\ \hline
 $0.1745_L$ & 2.460(55) & 0.5706(76) & 0.7927(43) & 0.8729(31) & 1.4350(94) &
 0.0544(10)
 \\ \hline
 $0.1746_L$ & 2.489(54) & 0.5616(47) & 0.7891(33) & 0.8700(19) & 1.433(23)  &
 0.05205(81)
 \\ \hline
 $0.1747_L$ & 2.457(48) & 0.5303(74) & 0.7566(75) & 0.8566(38) & 1.403(16)  &
 0.04602(77)
 \\ \hline
 $0.1745_H$ & 3.840(81) & 0.3991(86) & 1.0635(84) & 0.8232(27) & 1.096(16)  &
 -0.0260(15)
 \\ \hline
 $0.1746_H$ & 3.85(11)  & 0.481(11)  & 0.881(48)  & 0.8395(22) & 1.055(37)  &
 -0.0419(15)
 \\ \hline
 $0.1747_H$ & 3.98(11)  & 0.456(13)  & 0.996(36)  & 0.8375(26) & 1.028(42)  &
 -0.0403(23)
 \\ \hline
 $0.1750_H$ & 3.884(91) & 0.531(18)  & 1.0936(97) & 0.8690(46) & 1.064(37)  &
 -0.0525(24)
 \\ \hline
 $0.1755_H$ & 4.02(10)  & 0.7012(97) & 1.1056(99) & 0.9186(27) & 1.219(41)  &
 -0.0868(17)
 \\ \hline
\end{tabular}
\end{center}
\end{table}

\clearpage
\begin{table}
\begin{center}
\parbox{0.8\linewidth}{\caption{\label{tab3}\em
 Selected results of the runs on a $16^3 \cdot 32$ lattice at
 $\beta=3.35,\; a\mu_l=0.0075,\; a\mu_\sigma=0.2363,\;
 a\mu_\delta=0.2138$.
}}
\end{center}
\begin{center}
\renewcommand{\arraystretch}{1.2}
\begin{tabular}{*{7}{|l}|}
  \hline
  $\kappa_l=\kappa_h$ & \,\,\,\,\,\,\, 
  $r_0/a$ & \,\,\,\, 
  $am_{\pi}$ &\,\,\,\, 
  $am_{\rho}$ & \,\,\,\, 
  $am_{K}$ & \,\,\,\, 
  $am_{D}$ & \,\,\,\,
  $am_{\chi l}^{PCAC}$
 \\ \hline
 $0.1690$ & 2.222(54)  & 0.8237(13) & 0.9684(20) & 0.9231(11) & 1.3192(87) &
 0.12113(40)
 \\ \hline
 $0.1695$ & 2.503(41)  & 0.7329(11) & 0.8916(15) & 0.8652(11) & 1.2827(58) &
 0.09738(34)
 \\ \hline
 $0.1700$ & 2.812(48)  & 0.5857(18) & 0.7631(35) & 0.7739(12) & 1.223(23)  &
 0.06417(44)
 \\ \hline
 $0.1702$ & 2.87(16)   & 0.5082(26) & 0.7038(39) & 0.7379(22) & 1.187(21)  &
 0.04837(30)
 \\ \hline
 $0.1704$ & 3.28(12)   & 0.3695(22) & 0.6041(44) & 0.6553(21) & 1.110(31)  &
 0.02569(55)
 \\ \hline
 $0.1705$ & 3.31(13)   & 0.3433(52) & 0.5913(83) & 0.6480(18) & 1.080(35)  &
 0.02117(53)
 \\ \hline
 $0.1706$ & 4.50(20)   & 0.4331(74) & 0.780(35)  & 0.6756(13) & 0.943(46)  &
 -0.0428(22)
 \\ \hline
 $0.1708$ & 4.378(37)  & 0.4721(81) & 0.843(15)  & 0.7004(18) & 0.983(52)  &
 -0.0492(31)
 \\ \hline
 $0.1710$ & 4.59(16)   & 0.508(11)  & 0.812(16)  & 0.7216(17) & 0.957(20)  &
 -0.0569(26)
 \\ \hline
\end{tabular}
\end{center}
\end{table}

\clearpage

\clearpage

\clearpage

\clearpage
\begin{figure}
\begin{center}
\vspace*{0.10\vsize}
\begin{minipage}[c]{1.0\linewidth}
\hspace{0.06\hsize}
\includegraphics[angle=-90,width=0.80\hsize]
 {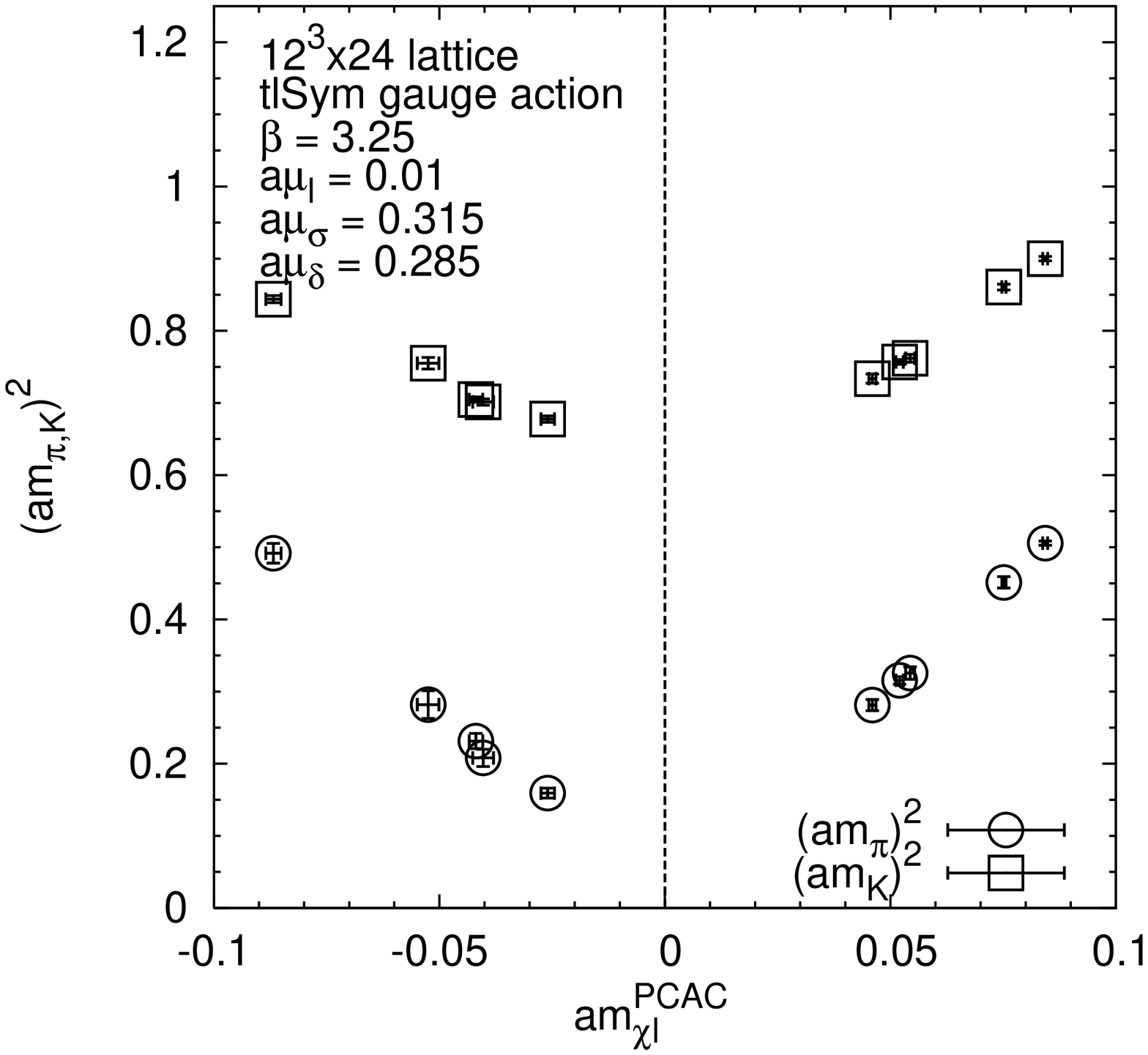}
\end{minipage}
\end{center}
\vspace*{-2em}
\begin{center}
\parbox{0.8\linewidth}{\caption{\label{fig_b325_amChi}\em
 The mass-squared of pion and kaon as a function of the untwisted
 PCAC quark mass on a $12^3 \cdot 24$ lattice at
 $\beta=3.25,\;a\mu_l=0.01,\;a\mu_\sigma=0.315,\;a\mu_\delta=0.285$.
 }}
\end{center}
\end{figure}

\clearpage
\begin{figure}
\begin{center}
\vspace*{0.10\vsize}
\begin{minipage}[c]{1.0\linewidth}
\hspace{0.06\hsize}
\includegraphics[angle=-90,width=0.80\hsize]
 {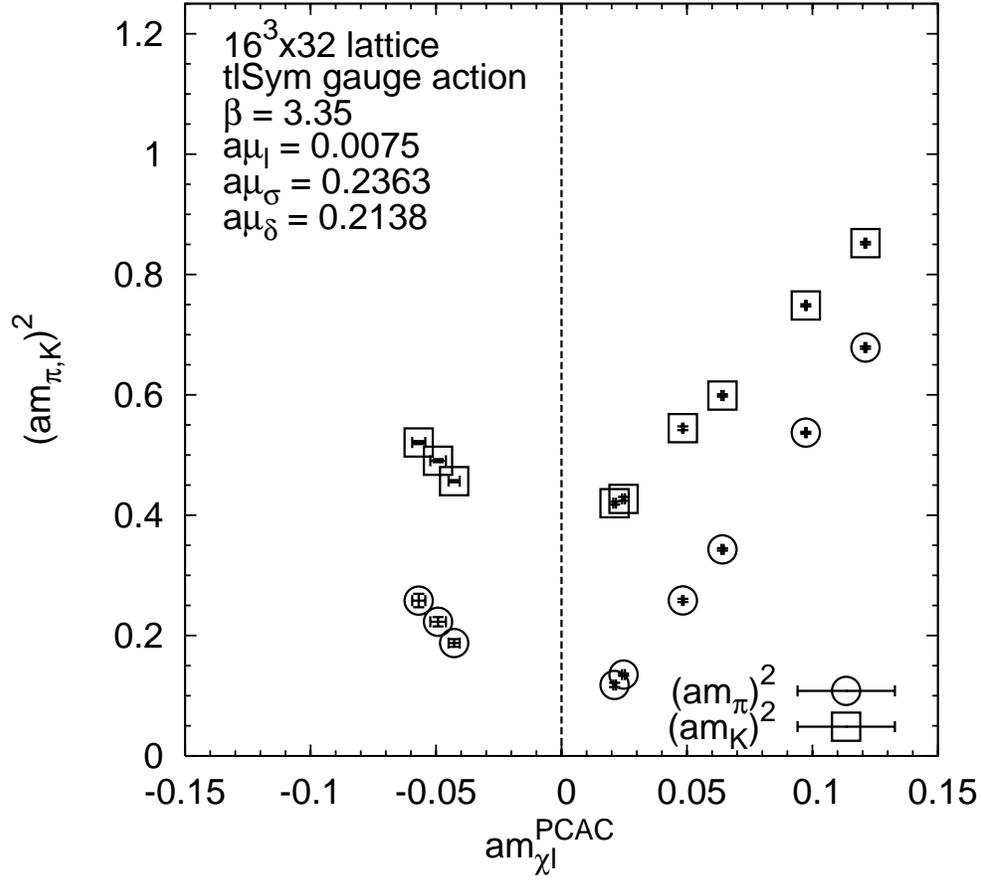}
\end{minipage}
\end{center}
\vspace*{-2em}
\begin{center}
\parbox{0.8\linewidth}{\caption{\label{fig_b335_amChi}\em
 The mass-squared of pion and kaon as a function of the untwisted
 PCAC quark mass on a $16^3 \cdot 32$ lattice at
 $\beta=3.35,\;a\mu_l=0.0075,\;a\mu_\sigma=0.2363,\;a\mu_\delta=0.2138$.
 }}
\end{center}
\end{figure}

\clearpage

\end{document}